# Single-molecule FRET dynamics of molecular motors in an ABEL Trap

Maria Dienerowitz[1*], Jamieson A. L. Howard[2,3], Steven D. Quinn[2,4], Frank Dienerowitz[5] and Mark C. Leake[2,3,4]

[1] Single-Molecule Microscopy Group, Universitätsklinikum Jena, Nonnenplan 2 – 4, 07743 Jena, Germany.
[2] Department of Physics, University of York, Heslington, York, UK. YO10 5DD.
[3] Department of Biology, University of York, Heslington, York, UK. YO10 5DD.
[4] York Biomedical Research Institute, University of York, Heslington, York, UK. YO10 5DD.
[5] Ernst-Abbe-Hochschule Jena, University of Applied Sciences, Carl-Zeiss-Promenade 2, 07745 Jena, Germany
* corresponding author: maria.dienerowitz@york.ac.uk

Abstract

Single-molecule Förster resonance energy transfer (smFRET) of molecular motors provides transformative insights into their dynamics and conformational changes both at high temporal and spatial resolution simultaneously. However, a key challenge of such FRET investigations is to observe a molecule in action for long enough without restricting its natural function. The Anti-Brownian ELectrokinetic Trap (ABEL trap) sets out to combine smFRET with molecular confinement to enable observation times of up to several seconds while removing any requirement of tethered surface attachment of the molecule in question. In addition, the ABEL trap's inherent ability to selectively capture FRET active molecules accelerates the data acquisition process. In this work we exemplify the capabilities of the ABEL trap in performing extended timescale smFRET measurements on the molecular motor Rep, which is crucial for removing protein blocks ahead of the advancing DNA replication machinery and for restarting stalled DNA replication. We are able to monitor single Rep molecules up to 6 s with sub-millisecond time resolution capturing multiple conformational switching events during the observation time. Here we provide a step-by-step guide for the rational design, construction and implementation of the ABEL trap for smFRET detection of Rep *in vitro*. We include details of how to model the electric potential at the trap site and use Hidden Markov analysis of the smFRET trajectories.

1. Introduction

The field of single-molecule biophysics is at the forefront of modern scientific tools, enabling the very building blocks of human life to be characterized with detail far surpassing those obtained from ensemble techniques [1]. While fast and efficient, ensemble measurements always generate an average picture: transient intermediate conformations and dynamics are often hidden. Extracting meaningful data from single molecules in complex liquid environments is inherently challenging yet offers the reward of accessing molecular dynamics, protein function and molecular heterogeneity [2, 3]. Fluorescence microscopy tools are proving invaluable for extracting details at the molecular length scale [4], with confocal microscopy and total internal reflection (TIRF) microscopy in combination with Förster energy transfer (FRET) currently being the most established techniques [5].

In biomolecular FRET measurements, the efficiency, $E_{FRET}$, of the non-radiative transfer of energy between an excited donor fluorophore and an acceptor, quantitatively reports on their separation distance, R, typically over the range 1-10 nm [6]. Given $E_{FRET} = 1/(1 + (R/R_0)^6)$, where $R_0$ is the distance between the fluorophores when $E_{FRET} = 0.5$, real-time fluctuations in the measured FRET efficiency can be correlated to conformational dynamics within single fluorescently-labelled biomolecules [7]. A major hallmark of the FRET approach is the ability to unveil population distributions of inter-dye distances, and reveal transient conformations that may be otherwise hidden by conventional ensemble-averaging. Coupled with the ability to probe individual molecules for long periods of time (seconds to minutes) with millisecond time resolution, the FRET approach has been successfully applied to reveal conformational dynamics across a wide range of proteins [8], nucleic acids [9-11] and biomolecular complexes [12-14]. However, one has to make a choice between investigating freely diffusing molecules for inherently short observation times or extend this observation time at the price of potential impairment in biological function by attaching molecules to a surface. Attachment assays work for some systems but finding the combination of FRET labels and an attachment strategy that do not impede the function of the molecule in question is a non-trivial enterprise[15, 16].

To address these limitations and observe proteins and molecular motors in action a growing field of research devoted to the development of single-molecule confinement tools gained momentum during the past decade [17-19]. The challenge is to monitor a single protein - typically 5-10 nm in effective diameter - long enough to detect conformational changes during one reaction cycle of its natural function while avoiding the need for surface immobilization. A protein diffuses 10 times faster than a typical virus, for example, because the rate of its Brownian diffusion is inversely proportional to its hydrodynamic radius as modelled by the Stokes-Einstein relation. In addition, single molecules are commonly labelled with one fluorophore only, thus giving out a limited number of photons to collect. Therefore, the key challenges are to 1) detect a fast-moving molecule with sub-millisecond time resolution or less, 2) hold it in place confined to a region of space whose effective length scale is sub-micrometre, and 3) record conformation changes as well as other molecular properties. Currently, the most promising single-molecule traps available are the Anti-Brownian ELectrokinetic Trap (ABEL trap) [20-23], geometry induced electrostatic fluid trap [24] and self-induced back action optical trap [25]. Only the first two techniques have demonstrated holding molecules as small as single fluorophores in solution.

In addition to confining a single molecule to an observation region, the ABEL trap provides access to photophysical properties of fluorescent labels and facilitates hydrodynamic profiling of individual molecules [26-28]. Measuring a molecule's diffusion constant *D* and electrokinetic mobility µ allows us to observe the binding of individual molecules to DNA [20, 23], for example. Recording fluorescent brightness, lifetime, anisotropy and spectral information has enabled the study of light induced conformational changes and oligomerisation effects of the photosynthetic antenna protein allophycocyanin [28, 29], the pigment-protein antenna complex C-phycocyanin [30] as well as discriminating between single- and double stranded DNA molecules in a mixture [31]. The ABEL trap has proven an excellent tool to investigate redox cycling in the multi-copper enzyme bNiR [32] and ATP hydrolysis in the multi-subunit enzyme TRiC [33] via fluorescence intensity changes. Combining single-molecule FRET and ABEL trapping offers a new approach to access the dynamics of conformation changes associated with single molecules at near shot-noise limited precision [34]. This is particularly promising for monitoring the complex multi-subunit dynamics of single molecular motors.

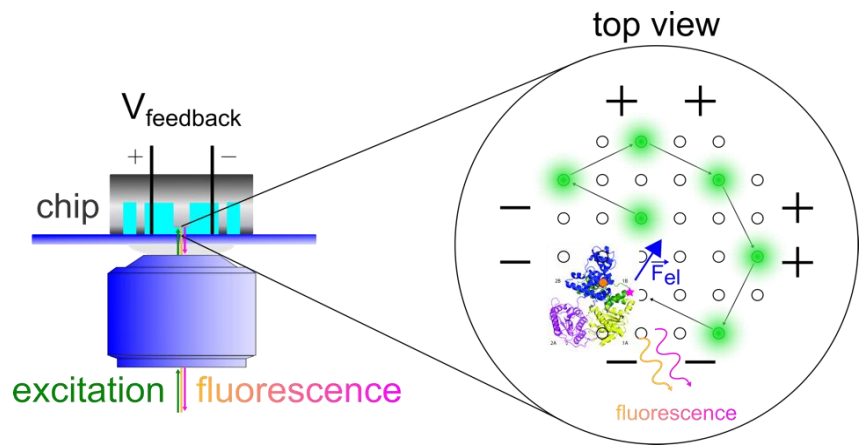

**Figure 1. Schematic of the ABEL trap.** The excitation laser (green dot) scans a predefined pattern within the trapping region. The fluorescence emitted by a trapped Rep molecule in a confocal ABEL trap setup provides the molecule's FRET signal and position information. This is the basis for calculating the feedback voltage required to generate the electrophoretic force $\vec{F}_{el}$ that pushes the molecule towards the centre of the trap.

Here, we probe the conformational dynamics of single Rep molecules in a home-built ABEL trap (Figure 1) (system developed previously in the lab of Michael Börsch, Universitätsklinikum Jena, Germany) [35-37]. Helicase molecular motors are essential in every aspect of nucleic acid metabolism, hydrolysing nucleoside triphosphates (NTP) (usually ATP) in order to generate energy for translocation along a nucleic acid strand in a directional manner [38-40]. Translocation generally occurs in one to two base steps per hydrolysis of each nucleotide triphosphate, disrupting the hydrogen bonding between complimentary bases at each step [41, 42]. *In vivo* nucleic acids are bound by proteins that must simultaneously be removed at the time of strand separation [43], making the process far more complex than simply disrupting base pairing. Helicases are able to utilise the excess free energy derived from NTP hydrolysis and base pair separation to produce force in order to push proteins along DNA [44-47]. It has been generally assumed that proteins are displaced from DNA in this manner, however the actual method of protein displacement still remains to be elucidated. In order to perform translocation, helicases undergo molecular conformational changes, and so are good experimental samples for optimising biophysical techniques that are designed to explore the switching between different molecular states.

Accessory replicative helicases are required in cells in order to facilitate the progression of the bacterial replication machinery called the replisome [48] through nucleoprotein complexes that would otherwise block replication [49-55]. Rep is the accessory replicative helicase in *Escherichia coli* bacteria that removes protein blocks to DNA replication bound to DNA as well as helping to restart the stalled replication machinery [56]. It contains four subdomains (1A, 2A, 1B and 2B), of which 2B can rotate around a hinge region allowing for a significant conformational change in the protein [39, 40, 57-62] (see Figure 2). As a superfamily 1A helicase, Rep has a 3'-5' directionality and as such likely displaces nucleoprotein complexes ahead of the advancing replication fork on the leading strand DNA template [49, 51].

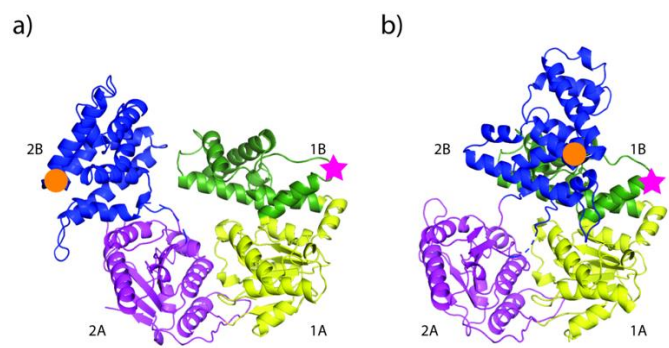

**Figure 2. The Rep Helicase Crystal Structures.** Rep contains 4 subdomains: 1A (yellow), 1B (green), 2A (magenta), and 2B (blue). In its crystal structures Rep can be seen to occupy both an a) open and b) closed conformation in which the 2B subdomain has rotated through approximately 130 ° [60]. The location of the attached dyes - Alexa Fluor 546 and Alexa Fluor 647 - used in this study are indicated by magenta stars and orange circles respectively. The placing of the dyes at positions A97C and A473C via maleimide linkage allow for the use of FRET as a nanoscale ruler for the measurement of distance between the residues on the 1B and 2B subdomains (PDB ID: 1UAA).

The key factors in identifying the efficacy of DNA replication are the frequency and duration of replisome pausing and stuttering. Accessing helicase motion and dynamics with FRET as a nanoscale ruler provides access to these quantities on a single molecule basis. In particular, labelling Rep at positions 97 (1B subdomain) and 473 (2B subdomain) with donor (e.g. Alexa Fluor 546) and acceptor (e.g. Alexa Fluor 647) fluorophores affords an opportunity to detect conformational transitions within the 2B subdomain (Figure 2). Positions 97 and 473 in Rep have been previously labelled both individually [63] or in tandem [61] and haven't impacted on protein functionality. These high-resolution measurements potentially expose transition rules of the Rep 2B subdomain stepping between open and closed states, the number of states and whether the presence of co-factors shifts the conformational state. To date, these types of measurements have required immobilization of the helicase to a substrate [61], that from our own experience risk impairing functional activity as well as introducing artificial photophysical effects due to the proximity to the surface. On the contrary, capturing Rep molecules in an ABEL trap allows for smFRET measurements away from any interfaces in solution for extended periods of time.

Here we describe a protocol for building and implementing the ABEL trap for identification of multiple conformational states in a molecular machine, exemplified by single Rep molecules using smFRET. We expect this procedure may open a platform for investigations of Rep mutations in particular as well as dynamic molecular machines and their interactions in solution in general.

2. Methods and Materials

2.1. Rep Purification and Labelling

The *rep* gene lacking all native cysteine residues and with non-native cysteines introduced in place of alanine 97 and alanine 473 was cloned into pET14b such that it was in frame with a hexa-histidine tag, creating the plasmid pJLH135. Rep was overexpressed and purified using pJLH135 as previously published for WT his-tagged Rep [64]. In short, the protein was overexpressed from *E. coli* using 0.2% arabinose at 20 °C for two hours before the resultant cell pellet was stored at -80 °C. The cell pellet was lysed using lysozyme and the polyethylene glycol hexadecyl ether Brij 58 before Polymin P and ammonium sulfate precipitations. The resulting supernatant was then purified using a 5 mL HisTrap FF crude column and a 3 mL heparin agarose column. The protein was then labelled using the following adapted version of a published method [65].

1) Collect fractions containing pure Rep as monitored by absorbance at 280 nm and SDS-PAGE and reduce for two hours at 4 °C by the addition of 5 mM Tris(2-carboxyethyl)phosphine hydrochloride (TCEP).
2) Add ammonium sulfate to 70% saturation whilst stirring at 4 °C. Continue stirring for a further 10 minutes.
3) Pellet the resulting precipitate by centrifugation at 18,000g for 20 minutes at room temperature. Resuspend the pellet in degassed labelling buffer (100 mM sodium phosphate pH 7.3, 500 mM sodium chloride, 20% (v/v) glycerol).
4) Alexa Fluor 546 $C_5$ Maleimide and Alexa Fluor 647 $C_2$ Maleimide (both Invitrogen A10258 and A20347 respectively) were dissolved in anhydrous Dimethyl sulfoxide (DMSO) and mixed in equimolar amounts. We note that while a wide range of organic dyes exist for labelling rep, the Alexa546 and Alexa 647 FRET pair displays lower hydrophobicity compared to spectrally similar Atto derivatives [66], and extends the FRET range beyond 10 nm, due to a Förster distance of 7.4 nm.
5) Add the mixed dyes to Rep at a 5-fold molar excess at mix by rocking at room temperature for 30 minutes. The maleimide groups present on the Alexa Fluor dyes bind covalently to the thiol groups (present on cysteine residues) in a highly selective manner. The labelling reaction is quenched by the addition of 10 mM 2-mercaptoethanol and rocking continued for a further 10 minutes.
6) Free dye is separated from labelled Rep using a 1 mL HisTrap FF crude column. The column is equilibrated with buffer A (20 mM Tris-HCl pH 7.9 and 500 mM NaCl, 20% (v/v) glycerol) + 5 mM imidazole before loading the dye/protein mix.
7) Wash the column with 20 mL of buffer A + 5 mM imidazole before developing it with a 20 mL gradient of buffer A + 5 mM imidazole to buffer A + 500 mM imidazole.
8) Collect the peak fractions containing labelled Rep and pool, aliquot and store at -80 °C in 20 mM Tris-HCl pH 7.9 and 500 mM NaCl, 30% (v/v) glycerol.
9) Final protein concentration is measured using the Bradford assay and dye concentration is measured using a NanoDrop 2000 (Thermo Scientific™). Labelling efficiency can be calculated as 100 * ([dyeA] + [dyeB]) / (2*[protein]). In this instance, the labelling efficiency was ~85 % with the extinction coefficients of Alexa Fluor 546 at 93,000 $cm^{-1}M^{-1}$ and Alexa Fluor 647 at 265,000 $cm^{-1}M^{-1}$.

All chemicals were purchased from Sigma Aldrich unless stated otherwise.

2.2 ABEL Trap

2.2.1 Basic Principle and Electric Field Modelling

The core principle of the ABEL trap relies on electrophoresis: generating an electrophoretic force $\vec{F_{el}} = q \cdot \vec{E}$ to displace a molecule with charge $q$ along a desired path by exposing it to a spatially uniform electric field $\vec{E}$. The total electrokinetic force is a combination of electrophoretic and electroosmotic forces acting on the surface charge of the molecule or the entire buffer solution respectively. In contrast to dielectrophoresis [67, 68], electrophoresis is a first order field effect and thus particularly suited for exerting forces on charged nanoscale objects. We simulate the electric field $\vec{E}$ for our specific chip geometry to assess its uniformity across the trapping region. This prompts the need for an electrodynamics simulation tool not easily available to everyone. We propose an alternative approach, exploiting the well-known mathematical analogy between electrostatics and steady-state thermal conduction [69]. Most finite element analysis tools include thermal conduction analysis as a standard feature implicitly providing the tools for electric field simulations.

| Electrostatics | Thermal conduction |
|---|---|
| Gauss's Law<br>$\Delta \varphi = 0$ | 1st Law of Thermodynamics<br>$\Delta T = 0$ |
| electric potential<br>$\varphi$ | Temperature<br>$T$ |
| electric displacement<br>$\vec{D} = -\varepsilon \cdot \nabla \varphi$ | heat flow<br>$\vec{h} = -k \cdot \nabla T$ |
| permittivity<br>$\varepsilon$ | thermal conductivity<br>$k$ |

**Table 1. Analogy between electrostatics and thermal conduction.**

The analogy between electrostatics and thermal conduction is straightforward, yet we stress that it is a mathematical and not a physical analogy. Table 1 outlines the analogy and provides an overview of how we substituted the electrostatic with thermal conduction variables. The electric potential $\varphi$ and the temperature field T correspond as Gauss's Law and the First Law of Thermodynamics state identical partial differential equations. The electric displacement $\vec{D}$ and the heat flow $\vec{h}$ yield the analogy between permittivity $\varepsilon$ and thermal conductivity $k$. In addition, the boundary conditions at material interfaces also have to be taken into account. In particular, the normal components of the electric displacement $\vec{D}$ and the electric potential $\varphi$ have to be continuous which we translate to continuous temperature T and normal components of the heat flow $\vec{h}$ across interfaces.

In the simulation tool the values of the electric potential $\varphi$ at the electrodes are to be set as temperature values T. The permittivity $\varepsilon$ of each domain is to be entered as thermal conductivity $k$. The values of the resulting temperature field T are in fact the values of the electric potential $\varphi$, which can be used to calculate the electric field

$$\vec{E} = -\Delta \varphi. \tag{1}$$

We demonstrate this method utilizing ANSYS Workbench for a 2D-horizontal plane through the trap (Figure 3). We use MATLAB (Mathworks) for post-processing of the data, i.e. calculating the electric field according to equation 1.

We adopted the following workflow:

1) Design geometry by specifying regions of electrodes, PDMS (Polydimethylsiloxane elastomer) and buffer solution.
2) Set boundary conditions: set temperature values at electrode positions which corresponds to applying voltages to electrodes.
3) Simulate the temperature field which corresponds to the electric potential.
4) Calculate the electric field according to $\vec{E} = -\nabla\varphi$ and derive the respective field lines in a post-processing step in MATLAB.

The results of the electric field simulation in our ABEL trap are presented in Figure 3 with typical feedback voltages of $\pm 1$ V and relative electric permittivities of $\varepsilon_r$=3 for PDMS, $\varepsilon_r$=80 for water and $\varepsilon_r$=1 for air. First, we consider the 1D case of applying a feedback voltage of +1 V at the left electrode and -1 V at the right electrode (Figure 3 a, b). This generates an electric field that is uniform in the central trapping region as the parallel field lines (white) show. In practice, the feedback voltage is applied to the x and y direction simultaneously with the resulting electric fields superimposing. An example is illustrated in Figure 3 c,d): we apply $\pm 1$ V across the x and y direction resulting in a diagonal uniform electric field (parallel field lines at the trap centre) suited to exert a force on a charged molecule in both x and y at the same time. Our analysis here can be compared to a more reduced model that was previously published in reference [36].

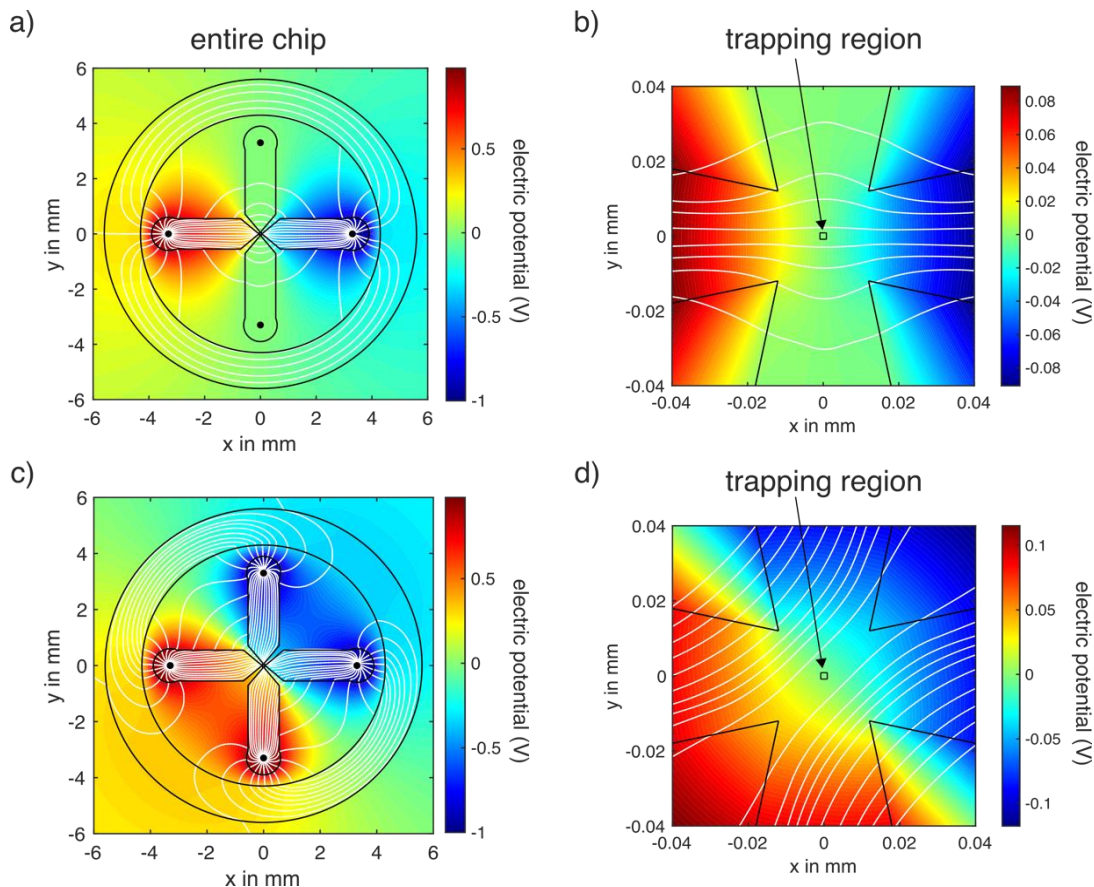

**Figure 3. Electric field simulation inside the ABEL Trap**. a) Feedback voltage is set to +1 V and -1 V across the x direction leading to a uniform electric field in the trapping area as demonstrated in a close up of the central region shown in b). c) The uniform electric fields along the x and y directions superimpose leading to diagonally oriented parallel field lines in the trapping area for applying ±1V across x and y simultaneously (zoom-in shown in d).

2.2.2 Feedback Trapping and Instrumentation

We implemented the ABEL trap in a confocal laser excitation volume smFRET setup by adding a custom-built laser scanning unit for beam steering and a microfluidics chip with platinum electrodes. The microfluidics chip plays a central role in holding the sample containing the Rep molecules in solution and delivering the electric field for ABEL trapping as outlined in Section 2.2.1. This prefabricated chip is made of quartz (fused silica) or a patterned PDMS (Polydimethylsiloxane elastomer) chip bonded to a microscope glass coverslip. The quartz chips have an excellent signal-to-noise ratio since the background from autofluorescence in quartz is significantly lower compared to standard glass coverslips. Quartz chips are relatively expensive to make and therefore are often cleaned and reused multiple times. Surface passivation by applying multilayer coating to prevent adsorption of the molecules to the quartz walls may be necessary [70].

The biggest advantage of the PDMS chips is the ease of handling. Their quick and cheap reproduction makes them disposable so there are no aggressive cleaning procedures to follow. More importantly, prior to the bonding process, the cover glass and the PDMS chip both reside in vacuum for plasma cleaning. This vacuum step removes any stored gases from the PDMS chip. The PDMS is then able to reabsorb air bubbles forming during filling the chip with the analyte. In this experiment, we used a chip made of patterned PDMS bonded on a standard microscope cover glass (24x32 mm, Roth, thickness #1). We purchased a customised mask wafer (Microfluidics Work Group, IPHT Jena) with the desired dimensions and patterns for 26 chips measuring 1.5 cm in diameter each. The protocols used for making the pattern design has been described and applied elsewhere [35-37, 71, 72]. We use a method for chip fabrication and handling similar to the one originated in previous publications [70, 73, 74] and measured the height of the central trapping area to be 600 nm [27].

At the heart of our setup is a field programmable gate array (FPGA 7852R, National Instruments): a real-time feedback control system running the laser scanning and calculating the required feedback voltages to trap a molecule according its position. We adapted previously published software to implement ABEL trapping capabilities on an FPGA with the LabVIEW FPGA module [20]. The excitation laser in an ABEL trap moves across a predefined pattern larger than the beam focus in order to determine the position of the molecule to trap. Because of the extent and speed of Brownian motion of proteins and objects smaller than ~100 nm, video-based camera tracking is too slow [22, 75]. A standard confocal system detects if there is a fluorescent molecule in the focal volume of the excitation beam but the resolution limit prevents us to know its exact position. Generating multiple detection sites spread out in space via laser scanning allows us to access a diffusing molecule's position while maintaining the excellent signal-to-noise ratio of a confocal system. Since the detected fluorescent photons and the execution of the laser displacement are time stamped, we are able to retrieve the molecule's position within the trap site (Figure 4).

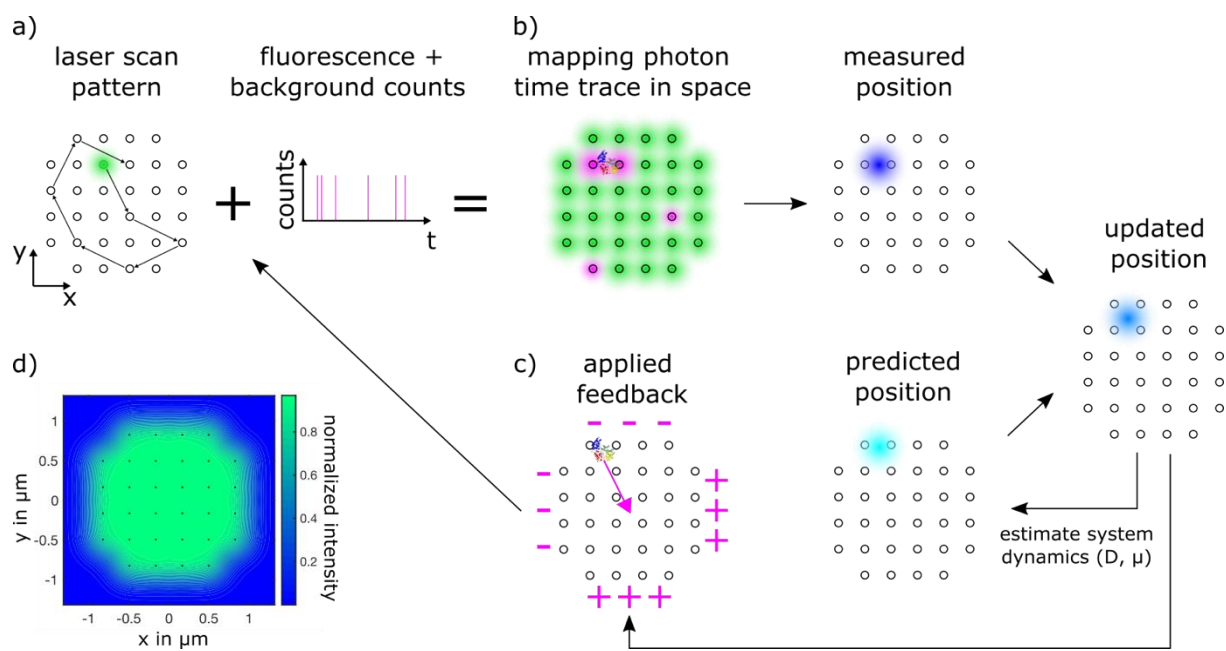

**Figure 4: Basic principle of the ABEL Trap.** a) An excitation laser scans a predefined 32 point pattern at 7 kHz (residing 4.5 µs at each point) while time stamped fluorescence photons originating from this scanning area provide position information of the molecule to trap. b) An FPGA controls the laser scanning, estimates the molecule's position with a Kalman filter including the transport parameters (D, µ) and c) sends out the required feedback voltages producing the electrokinetic force to trap the molecule at the centre of the scan pattern. d) A calculated time averaged intensity pattern of the entire trap site.

For the ABEL trap to work for single molecules, position detection based on the laser position alone is not accurate enough; the laser focus waist has a lateral diameter that is two orders of magnitude larger than a trapped molecule's typical effective diameter, and a molecule also continues to perform Brownian motion between scanning steps. Increasing the scanning speed is limited by the specifications of the beam steering unit (see section 2.2.3) and the signal-to-noise ratio. Decreasing the scan speed improves the signal-to-noise ratio up to the point where motion blur becomes apparent [76]. To achieve the required precision in position detection we implemented a Kalman filter [20, 77]: a standard tool for noisy measurement data in combination with known system dynamics - Brownian motion and electrokinetic response in our case. This filter estimates the position of the molecule in the ABEL trap based on the detected photons, the laser position and a prediction of the molecule's Brownian motion given its estimated diffusion coefficient $D$ and electrokinetic mobility $\mu$. The filter then weights the raw data position information (photon counts) with the position estimate based on the transport parameters ($D$ and $\mu$) to predict an updated position estimate (Figure 4c). To overcome a Kalman filter's limitations of fixed model parameters ($D$, $\mu$) it is possible to extend it using innovation whitening, including background noise with a maximum-likelihood estimator and thus achieve full real-time online determination of *D* and µ as demonstrated in [23, 72].

The details of our instrumentation are as follows (Figure 5):

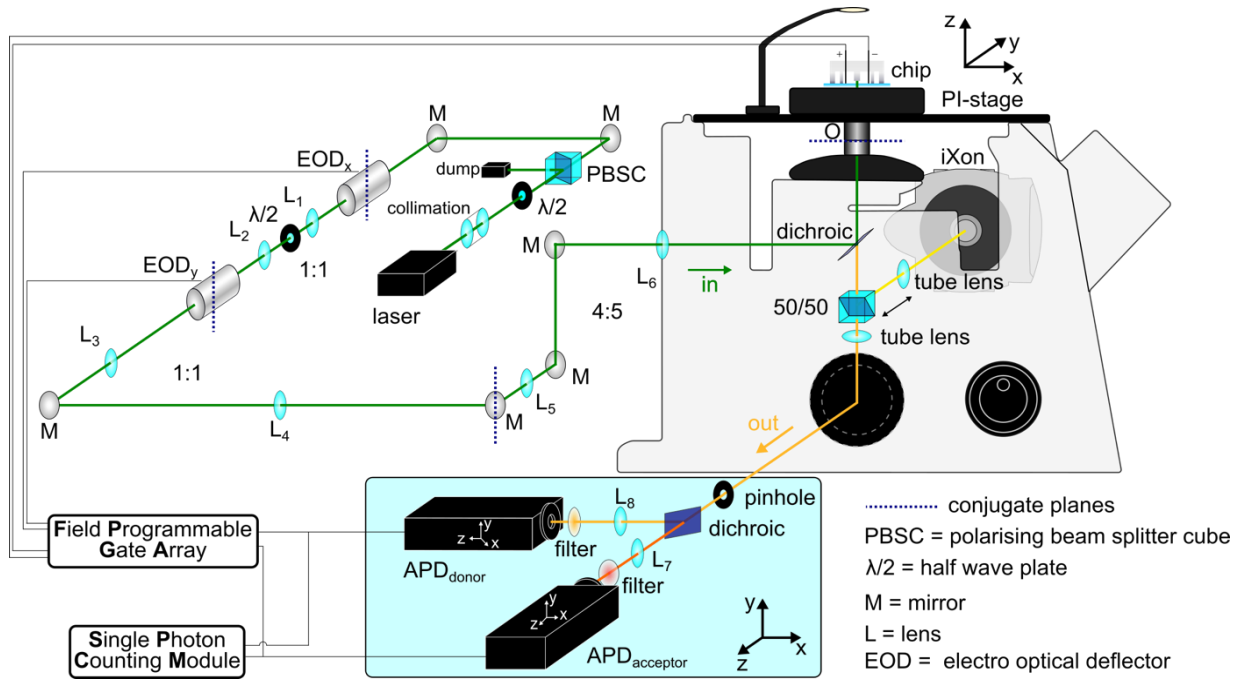

**Figure 5. Sketch of the light microscope setup with the implemented ABEL trap.** Telescopes ($L_1$-$L_6$) relay conjugate planes and adjust the beam diameter before entering the objective lens (O). The ratios given next to the telescopes indicate the focal length ratios of the telescope lenses relaying the conjugate planes of the laser beam path. Top left: laser excitation beam steering, bottom blue box: single photon fluorescence detection.

The linearly polarised 532 nm continuous excitation beam passes a half-wave plate and beamsplitter cube (AHWP05M-600, PBS201, Thorlabs) to adjust the laser power to 20 µW before entering the electro-optical deflectors (EOD, M310A, Conoptics) creating the scan pattern at the trap site. A dichroic beamsplitter (z 532 RD, AHF) inside a commercial inverted microscope body (IX71, Olympus) directs the excitation laser into the microscope objective lens (60x, oil immersion, PlanApo N, NA 1.42, UIS2, Olympus) while allowing the returning fluorescence to pass into the detection beam path at the same time. The 300 µm pinhole (P300D, Thorlabs) in the first image plane after the tube lens is larger compared to a standard confocal setup so as to not cut off any of the extended laser scanning pattern of the trap. A second dichroic beamsplitter (F48-640, AHF) spectrally separates the donor and acceptor photons towards their respective avalanche photodiode (APD SPCM-AQRH 14, Excilitas). After passing additional filters (donor bandpass: F37-582/75, acceptor longpass: F46-647, both AHF) a TCSPC (time correlated single photon counting) card (SPCM, DPC230, Becker&Hickl) and the FPGA (7852R, National Instruments) detects the donor and acceptor photons simultaneously. The TCSPC card time stamps the incoming photons for further time trace analysis. We extracted the recorded single molecule bursts using home-written software comprising an inbuilt C script in MATLAB (Mathworks).

The FPGA estimates the position of the trapped fluorescent molecule using a Kalman filter and applies appropriate feedback voltage to keep it in the trap. A feedback voltage amplifier ($V_{PP}$ = 20 V, 80 MHz operation frequency) connects the FPGA feedback output with the four platinum electrodes, delivering the electric field to trap the molecule into the microfluidics chip. A piezo stage (P-527.3CD, Physik Instrumente) holds the microfluidics chip and allows us precise axial position adjustment (spatial resolution 0.1 nm, repeatability ±1 nm along z) along the excitation laser propagation. Before starting an experiment, we utilize an EMCCD camera (iXon$^{EM+}$ DU-897, Andor Technology) to align the chip with respect to the trapping laser pattern by inserting a 50/50 beamsplitter into the beam path that is subsequently taken out of the detection beam path again. A standard halogen lamp serves as the illumination source.

2.2.3 Excitation Beam Alignment and Beam Steering

The beam steering in an ABEL trap demands for a fast and precise laser scanning method. While there are ample possibilities - from galvo mirrors to spatial light modulators - the speed requirements narrow our options down to two devices: electro-optical deflectors (EODs) and acousto-optical deflectors (AODs). AODs have larger deflection angles and are the more economical choice. They modulate the refractive index of a transparent crystal with acoustic waves (applying radiofrequencies (RF) to the crystal). The speed of the beam deflection depends on the beam diameter (smaller equals faster), using longitudinal (faster) or transverse waves and whether the RF synthesizer is addressed with an analogue voltage control oscillator or direct digital synthesis (faster). In addition, the diffraction efficiency decreases for larger angles (maximum modulation bandwidth) in practice up to 50% that has to be corrected for homogenous illumination of the entire trapping pattern[70].

Electro-optical deflectors (EODs) are faster than AODs and operate based on the Pockels effect that changes the optical properties of an electro-optical crystal in response to an applied electric field. The deflection angle is not wavelength dependent as in AODs, but a function of the refractive index dispersion and thus constant over the visible wavelength range. EODs deflect the entire beam with the transmission efficiency independent of the deflection angle and they allow the beam to pass straight through when switched off. These properties make setting up multiple excitation wavelengths easier with EODs. Although they generally have a smaller deflection angle compared to AODs, there is no decrease in intensity for larger deflection angles, making a correction for a homogeneous excitation pattern redundant.

We use a pair of coupled EODs (M310A, Conoptics) in combination with wideband power amplifiers (7602M, Krohn-Hite Corporation) delivering the required high voltage (±200 V). The stated 1.5 µrad/V deflection angle with an aperture of 2.5 mm allows us to generate a pattern of 2.3 µm x 2.3 µm with 6 resolvable spots along the x and y direction in our sample plane. In this experiment we trap with EODs at 7 kHz. Faster scanning requires better signal-to-noise ratios to be able to detect the molecule's position accurately enough to trap it. Once the decision for a beam steering system is made, the only free parameter determining the size of the pattern and number of resolvable spots is the beam diameter. Decreasing the laser beam diameter results in a larger diffraction angle but limits the number of spots resolvable and increases the beam waist in the sample plane. The best way to determine the required input beam diameter is to start with the beam waist and Rayleigh range at the sample plane and work our way back to the laser source.

Figure 4a) shows the intended scan pattern: a Knight's tour pattern with 32 points, diagonally separated by 0.47 µm. The Knight's tour pattern can complete a closed tour of all points (ending and starting at the same point) while visiting the entire trapping area throughout one scan (as opposed to a snake scan that starts at one side and ends on the opposite side) and was first demonstrated by Wang et al. [78]. In order to create a homogeneous excitation intensity across the entire trapping area we aim for a beam waist of $\omega_b$=0.5 µm. The ABEL trap actively confines the molecule's Brownian motion along x and y, restricting its z movement within the physical limits of the microfluidic chip. Therefore, the confocal parameter b of the beam has to remain constant across the entire chip height to obtain the same fluorescence signal irrespective of the molecule's actual z position along the excitation beam. This condition is fulfilled with the confocal parameter b=2.95 µm obtained by doubling the Rayleigh range ($b=2\pi\omega_b^2/\lambda$).

To determine the beam diameter d=2$\omega_i$ entering our microscope objective we approximate Gaussian beam propagation through a lens for small angles with [79]

$$\omega_i = \frac{\lambda \cdot f_{obj}}{\pi \cdot \omega_b}$$

The objective lens focal length is calculated with $f_{obj}$ = $f_{tubelens}$ / $\mathcal{M}$ (with the objective lens magnification $\mathcal{M}$), in our case $f_{obj}$ = 3.33 mm, that leads to a required beam diameter of d=1.13 mm (note, this reciprocal relationship between input beam width and output width at the sample is also utilised in single-molecule Slimfield microscopy [80]). The above equation for collimating a Gaussian beam is only valid if $\omega_i$ >> $\omega_b$ and if $\omega_i$ >> $\lambda$ – the requirement for the paraxial approximation. Although these equations only give estimates, they are usually sufficient and the beam size can be adjusted easily with an appropriate Keplerian telescope. The most important aspects of using telescopes is to relay conjugate planes. First, this is essential for mapping the x deflection of the beam onto the y deflection and further into the trapping plane to ensure the beam manipulation along both axes is in the same plane creating the 2D scanning pattern (lenses $L_1$ and $L_2$ in Figure 5). Second, we need to ensure that the beam always enters the objective's back aperture. By relaying the conjugate planes of x and y deflection onto the objective's back aperture only the angle of the beam entering the objective changes but not its spatial position. The objective's back aperture is the Fourier plane of our sample plane translating the angular deflection into lateral movement. The additional telescope in our beam path (lenses $L_3$ and $L_4$ in Figure 5) helps with aligning the entire system. This gives us another conjugate plane positioned directly on a mirror in order to accurately align the beam focus without having to realign the rest of the beam path. The last telescope (lenses $L_5$ and $L_6$ in Figure 5) relays the conjugate plane of this beam steering mirror on the back aperture of the objective.

We adopted the following procedure to calibrate the EOD scan pattern:

1) Set EOD deflection scale to start value. Make sure there is no gain induced voltage clipping of the power amplifier gain.
2) Set a 4-point square scan pattern and measure their distance at different pre-defined spacings with a camera. The camera has to be calibrated too to infer spacings from the pixel distance.

3) Compare set distance with measured distance and adjust EOD deflection scale accordingly. Re-iterate until set distance and measured distance match.

Aligning avalanche photodiodes (APDs) detection paths

The donor and acceptor APD are both attached to one x-y-z micrometre translation stage including filters, lenses, dichroic beam splitter and pinhole. Both detectors have additional freedom of micrometre movement along x and y with respect to the beam propagation axis.

1) Place an autofluorescent plastic slide (Part No: 92001, Chroma) on microscope objective for alignment after adding immersion oil.
2) Align APD in straight beam path first (acceptor APD in our case) by repeatedly adjusting x,y,z positions of the main translation stage positioning the pinhole correctly in the detection beam path followed by fine adjustment in x and y of the APD position.
3) After maximising photon count rate on the acceptor APD repeat the same for donor APD taking care not to misalign the pinhole.
4) The z position of the pinhole as well as the APD is a little less critical in our case since the beam waist of the excitation laser is slightly extended and the pinhole is large compared to a confocal setup. Final z position adjustment of the APDs is best conducted with a working FRET probe – the one we intend to experiment with.
5) Most crucial for a signal-to-noise ratio of at least 2:1 and the ability to trap is the correct axial positioning of the microfluidic chip that we accomplish with the piezo stage (P-527.3CD, Physik Instrumente) holding the sample stage.

3. FRET data acquisition: A general strategy to identify conformational fluctuations in Rep.

A 10 µL solution containing 150 pM labelled Rep (Alexa Fluor 546, Alexa Fluor 647) was prepared and injected immediately into a pre-fabricated ABEL chip. We adjusted the protein concentration in a way that it is dilute enough to avoid multiple molecules entering the trap site while trapping one molecule and dense enough that waiting time between molecules entering the trap site is reduced to 10 s. 150 pM was sufficient in order to observe ca. 1-5 simultaneous single-molecule bursts of donor and acceptor fluorescence emission over a 10 second timescale, as indicated in Figure 6a), reflecting the presence of single freely diffusing doubly-labelled rep species across the confocal volume when the trap was turned off. The buffer conditions (10 mM Tricine, 10mM succinate, 1.25 mM $MgCl_2$, 0.03 mM KCl, 10 mM NaCl, pH 8) were chosen to facilitate ABEL trapping whilst also enabling conformational freedom between high and low FRET states. All experiments were performed in a temperature-controlled laboratory at 21 °C.

In the absence of feedback, single Rep molecules diffuse through the laser scan pattern with a typical diffusion time of around 30-40 ms (Figure 6a). With feedback applied however, Rep molecules that diffused into the trap were rapidly pushed towards the centre and held for approximately half a second, prior to photobleaching or escaping the trap (Figure 6b). Trapping based on the acceptor fluorescence signal helps to maximise capturing FRET active molecules instead of donor-only molecules or aggregates. As highlighted previously, the mean trapping time varies as a function of excitation power, demonstrating the complex interplay between diffusional escape from the trap, photobleaching or transitions to long-lived and non-emissive dark states. In this work, we observed a typical trapping time of 475 ms at 20 µW (Figure 6b), even in the absence of oxygen scavengers. Although a direct comparison is not straightforward, owing to variations in biomolecular diffusion time, this was comparable to those observed from trapped Alexa647-labelled dsDNA molecules under similar working conditions [20]. Since the ABEL trap is an active trap, a second molecule entering the trap site would result in both escaping. We noted that >10% of all trapped Rep molecules resided in the trapping region for >1s (Figure 6c) and recorded an average of approximately 27,000 donor and acceptor photons per 1s trapping event. We extract the FRET time traces from the donor and acceptor photon counts as obtained from the APD detection. A single photon counting card (DPC-230, Becker&Hickl) records the photons' absolute time of occurrence at a maximum time resolution of 165ps. We chose to present the so acquired photon data with a 1ms integration time.

Typical signal-to-noise ratios varied between 2:1 and 3:1 during a trapping event. A measurement time window of 30 minutes allowed us to collect on average 300-500 trapping events. The illumination was uniform across the trapping region (see Figure 4d) and thus fluctuations in emission intensity were ruled out from arising from residual molecular motion. As can be seen from the representative traces shown in Figures 6d) and 8a), anticorrelations between the donor and acceptor emission fluorescence intensities gave rise to fluctuations in the apparent FRET efficiency $E_{FRET}$. We note that every trapping event ended with a step to background fluorescence, indicating either photobleaching or the diffusion of the trapped molecule out of the trap. Occasionally, short-lived intensity bursts were observed indicating the approach of a second Rep molecule within the vicinity of the trap. Aggregates exhibited a distinctly different photon trace with stepwise photobleaching and the maximum count rates far exceeding the ones of the single molecule trapping events, which remained at a constant level as we illustrate in Figure 6b) throughout the experiment with the same microfluidic chip.

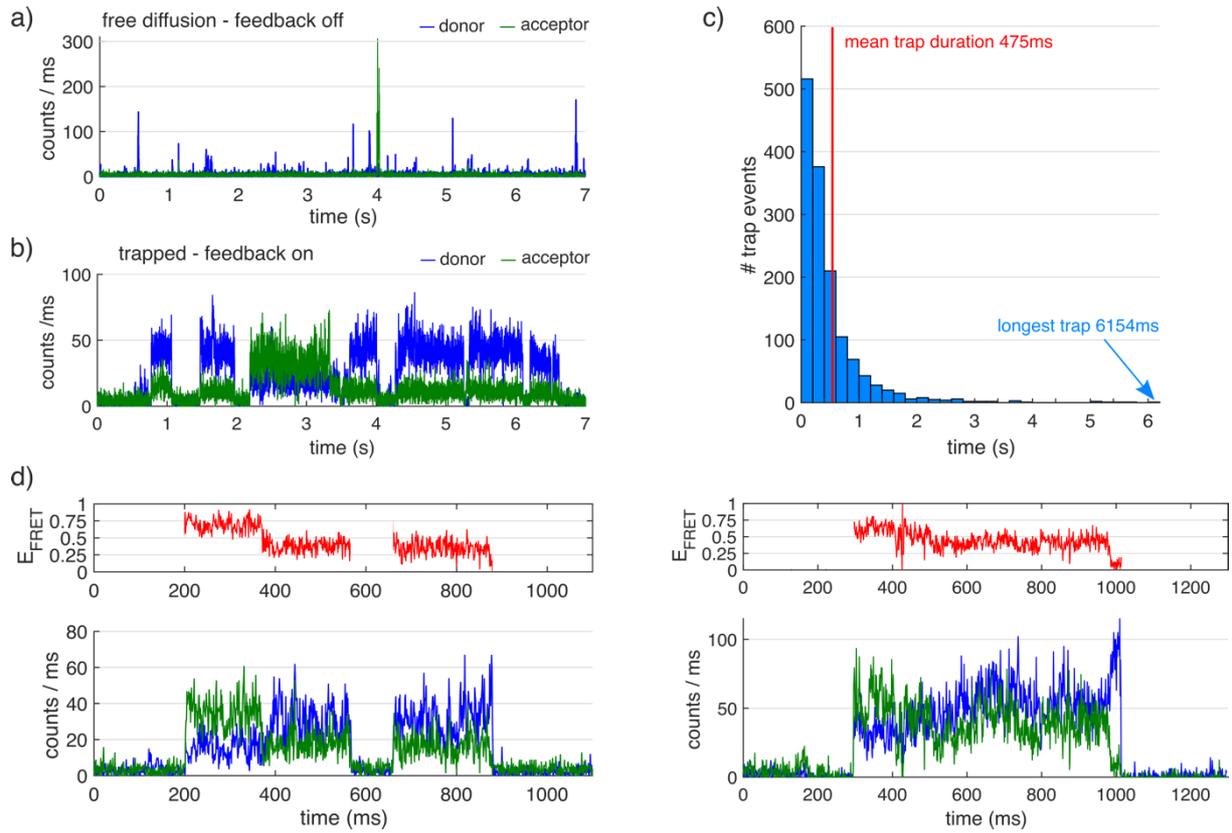

**Figure 6. ABEL trapping of doubly-labelled Rep molecules in solution.** (a) In the absence of applied feedback, the donor and acceptor fluorescence time traces showed brief bursts of typically a few tens of ms in duration above background. (b) In the presence of feedback, the molecular residence time lasted typically half a second, but, sometimes up to several seconds. (c) Histogram of recorded trap durations. (d) Trapping of Rep showed anti-correlations between donor (blue) and acceptor (green) fluorescence trajectories, corresponding to observable fluctuations in FRET efficiency (red).

We determined the FRET efficiency $E_{FRET}$ via $E_{FRET} = \frac{N_A}{N_A + N_D}$ where $N_D = N_{D,measured} - B_D$ and $N_A = N_{A,measured} - B_A - \alpha N_D$. $N_D$ and $N_A$ represent the total counts recovered from the donor and acceptor channel, $B_D$ and $B_A$ are the respective background counts and $\alpha$ is the leakage fraction of donor emission into the acceptor channel, as determined by analysis of trajectories displaying donor-only signals and recovered via $\alpha = avg\left(\frac{N_A}{N_D}\right)$. To estimate mean average background rates $B_D$ and $B_A$ for the donor and acceptor channels we fitted a Poisson distribution of the form $y = p_0 + \frac{e^{-B_{D/A}} B_{D/A}^x}{x!}$ to the probability density histogram from the background regions where $p_o$ is a constant of the fit. In our experiments, the recovered values were $B_D$=3.97 ± 0.04 (± S.D.) counts ms$^{-1}$, $B_A$=10.0 ± 0.05 counts ms$^{-1}$ and $\alpha$=0.075 (see Figure 7).

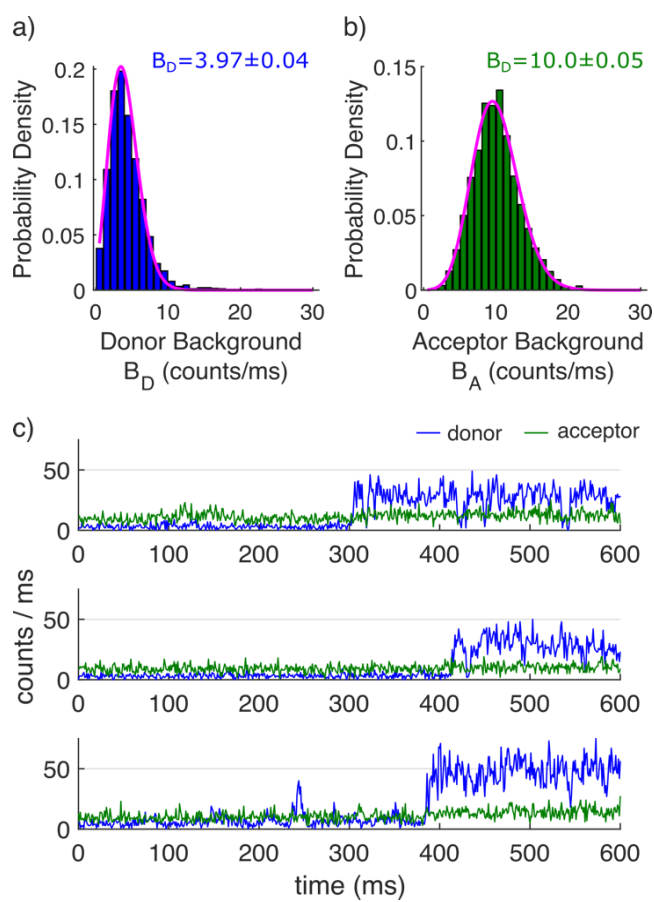

**Figure 7. Mean background rates for the donor and acceptor channels.** (a) Histograms and Poisson fits (solid line magenta) associated with Alexa Fluor 546 (donor) and (b) Alexa Fluor 647 (acceptor) background rates. (c) Representative traces showing the trapping of Rep monomers containing donor-only (Alexa Fluor 546) and minimal bleed-through into the acceptor (Alexa Fluor 647) detection channel.

FRET time trajectories from single trapped Rep molecules record conformational fluctuations between open and closed forms - corresponding to low and high FRET states respectively - together with interconversion rates (see Figure 2 for open and closed conformation states). We fitted the FRET versus time trajectories displaying discrete quantized state fluctuations with a Hidden Markov Model (HMM)[81], a valuable high-throughput analytical approach for generating objective fits to large bodies of FRET data. The HMM model is an established technique for recovering the otherwise hidden idealized trajectory and has been applied extensively across the single-molecule community to identify protein-nucleic acid interactions[82], to analyse multi-chromophore photobleaching trajectories[83] and to differentiate between single-molecule FRET states. Briefly, the model utilizes the Viterbi algorithm to extract the most probable sequence of states for a given data set.

In this application, the FRET trajectories were modelled via the HaMMY algorithm[81] using a maximum of 10 different putative FRET states. Based on the convergence of the goodness-of-fit, we find that Rep adopts at least four rapidly interconverting conformational states under the conditions tested. We applied the method to all experimental FRET trajectories extracting the true number of states. Representative trajectories and reconstructed HMM fits are shown in Figure 8.

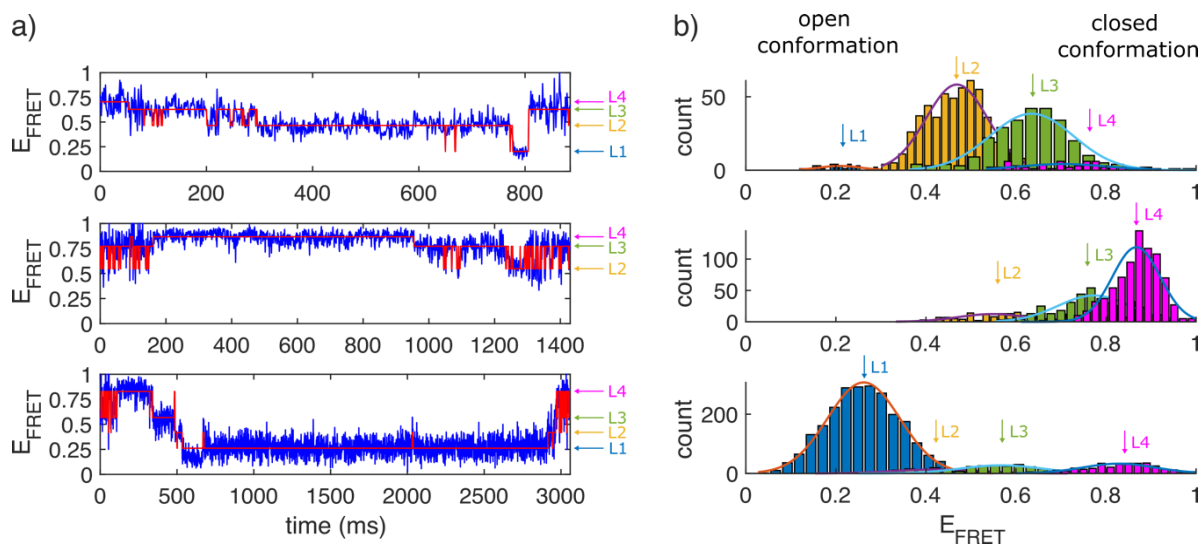

**Figure 8. Representative FRET traces and idealized trajectories obtained by Hidden Markov Modelling.** Representative trajectories (blue) and reconstructed HMM fits (red). b) corresponding FRET efficiency histograms.

While a major benefit of the FRET method applied to rep is to report rapid conformational fluctuations and transition kinetics via application of hidden Markov Modelling, it is important to note some of the limitations of the technique in this instance. First, not-trivial covalent tagging of the rep structure with appropriate donor and acceptor fluorophores is an essential requirement since intrinsically fluorescent amino acids within the structure, including tryptophan for example, are not sufficiently emissive or photostable for single-molecule detection. As such, care must be taken to ensure rep activity is maintained, regardless of the labelling position or choice of fluorophores. Second, the labelling method is stochastic, meaning under the current setup, any rep lacking a donor molecule goes unobserved, whereas if rep lacks an acceptor, a zero FRET population is recorded. Third, absolute distance measurement is challenging due to the dependence on likely fluctuations in the local environment and orientation of the fluorophores. Nevertheless, the data obtained by ABEL-FRET points towards the 2B subdomain of rep populating at least four conformational states as opposed to the two identified by X-ray crystallography (Figure 2), and holds exciting promise for unveiling the kinetic mechanisms underpinning helicase activity.

## 4. Conclusions and Outlook

The ABEL trap characterizes time-dependent changes in individual trapped molecules for extended periods of time without the need for surface functionalization for molecular immobilisation. Fluctuation techniques such as FCS, which only provides access to the several ms taken for a molecule to diffuse through the confocal volume, cannot resolve molecular processes unless they occur on a timescale faster than the diffusion time. Conventional widefield imaging setups such as TIRF with time resolutions of typically 10-50 ms struggle to access short-lived transitions and rely on the attachment of proteins to a surface. The ABEL trap thus opens the possibility for observing key molecular transitions that may be otherwise hidden. Its ability to measure intensity and FRET efficiency simultaneously also allows to identify aggregates and species to be distinguished based on reported FRET levels. This is important for the unambiguous dissection of hidden heterogeneity within the samples.

We expect this ABEL trap-based molecule-by-molecule analysis to be broadly useful in the context of investigating many different types of molecular machines in the presence of their physiological cofactors, as well as studying a plethora of dynamic molecular interactions, both conformational changes intramolecularly and intermolecularly.

5. Acknowledgements

We thank Michael Börsch, Universitätsklinikum Jena, for his support in providing unconstrained access to the ABEL trap instrumentation in his laboratory. We acknowledge support of the Physics of Life Group at the University of York, UK, the BBSRC (BB/P000746/1, BB/R001235/1), the Leverhulme Trust (RPG-2017-340) and Alzheimer's Research UK (ARUK-RF2019A-001).